\documentstyle[prb,aps,twocolumn,amssymb,amsfonts,epsfig]{revtex}

\begin{document}


\title{Temperature dependence of magnetism near defects in SrB$_6$.} 

\author{T. Jarlborg}

\address{
D\'epartement de Physique de la Mati\`ere Condens\'ee,
Universit\'e de Gen\`eve, 24 quai Ernest Ansermet,
CH-1211 Gen\`eve 4, Switzerland}

\date{\today}
\maketitle
\begin{abstract}
The T-dependence of magnetic moments in SrB$_6$ is studied through
spin-polarized band calculations for a supercell of Sr$_{27}$B$_{156}$ containing a B$_6$ vacancy. 
The magnetic moment decays rather quickly
with T despite the fact that only electronic Fermi-Dirac effects are included. This
result and the T-dependence of moments near a La impurity
can hardly explain the reports of a very high Curie temperature in hexaborides.
but suggest that the magnetism is caused by some other type of impurity.
\end{abstract}

\vspace{0.5cm}
Early reports of weak ferromagnetism (FM) in lightly doped CaB$_6$ or SrB$_6$, associated
with an unusually high Curie temperature (T$_C$) \cite{you}, have been followed by several theoretical
and experimental studies. For instance, it has been speculated that the FM could be
of excitonic origin \cite{ric}, be a spontaneous polarization of the dilute electron gas \cite{ort},
or be caused of special conditions around impurities and thereby lead to band magnetism \cite{tj,tj2}.
The band structure of the pure hexaboride \cite{mass}, which is metallic when
calculated within the density functional (DF) approach,
has been questioned, since one GW-calculation found a gap at the Fermi energy, $E_F$ \cite{tro}.
These results are in contrast to a recent independent GW-calculation by
Kino {\it et al}, where the ground state is metallic \cite{kino}.
 Also, there are conflicting results concerning the interpretation of the
experimental results on the weak magnetism. Quantum oscillations indicate that the FM
state is not intrinsic to the bulk, but that impurities might be important \cite{tera}. 
This conclusion was supported by Mori and Otani \cite{mori} who found that the FM signal could be removed by
a treatment of the crystals, and suggested that that the FM is due to iron impurities.
Other defects such as vacancies have been suggested as well \cite{fis}.
 Possible vacancies of whole B$_6$-clusters were anticipated in calculations
 by Monnier and Delley \cite{monnier}
who showed that a sizable FM moment surrounded such a vacancy in a
 supercell of CaB$_6$.  They suggest that the observation
of weak ferromagnetism in dilute doped hexaborides is caused either by a surface or interface effect,
since the B$_6$ vacancy is modeling the natural cleavage planes in polycrystals,
or by some amount of B$_6$ vacancies in pure crystals.

In the present letter, by use of the linear Muffin-Tin orbital (LMTO) 
band method in the local spin-density approximation,
we study the variation of the magnetic moment around a B$_6$ vacancy in Sr$_6$ as function of
temperature. This is done by self-consistent spin-polarized calculations, in which T-dependent
electronic excitations are modeled by the Fermi-Dirac distribution, but no
thermal disorder of the lattice is taken into account.  As in ref. \cite{monnier}
we consider a 3$x$3$x$3 supercell with one B$_6$ vacancy in SrB$_6$ so that the total
cell is Sr$_{27}$B$_{156}$. The missing B$_6$ atoms form empty MT-spheres, 
and the calculations include additional empty spheres
in the open part structure. The cell contains 19 nonequivalent MT-spheres, 4 Sr, 8 B and 7 empty ones.
 Other details of the calculation are as in ref. \cite{tj2}.
The density-of-states (DOS) functions near $E_F$ for the non-polarized
case, are shown in Fig. 1 for two sets of k-points (4 and 10) in the irreducible part of the Brillouin zone. 
It is seen that the essential features of the DOS are developed already when using 4 k-points,
with only small differences in the fine details. In particular, the DOS peak at $E_F$ is similar 
in the two sets
of k-points, so that the criterion for Stoner magnetism should come out well already when using
 only 4 k-points. 
The Stoner factor
is calculated to be 1.05, just above the limit for ferromagnetism, 1.0, indicated by the horizontal 
line in fig. 1. The broken line shows the partial DOS from the six equivalent B-atoms closest
to the B$_6$-vacancy. It is seen that roughly 1/3 of the total DOS at $E_F$ comes from these sites
although they represent only about 3 percent of the total number of atoms in the cell.
 In the spin-polarized calculations (at low T),
when using 4 k-points,
one finds a magnetic moment, $m=$2.1 $\mu_B$ per cell, in agreement with ref. \cite{monnier}. 
This moment decreases to
about 1.8 $\mu_B$ when the number 
of k-points is increased to 10 for the same temperature. 
This type of magnetism is a standard Stoner magnetism \cite{c15}, where the large DOS
on the 6 B-atoms nearest to the vacancy contribute most to the Stoner instability, cf. Fig 1.
 Similarly,
about 40 percent of the total moment in the spin-polarized calculations comes from  
these six B-sites closest to the vacancy. 

Spin-polarized calculations at different temperatures in the Fermi-Dirac
distribution, show that the moment goes to zero when T is increased to about 700 K. The two
sets of k-points have been used at the lowest and highest T, and both sets give a vanishing moment
at T=700K. Calculations for intermediate T, using 4 k-points,
show first a gradual decrease of $m$ at low T, while the curve drops sharply as T is approaching 700 K,
as shown in Fig 2.
This T-dependence is understood from the fact that $E_F$ falls on a narrow peak in the DOS, so that  
thermal smearing, due
to electronic excitations, 
can reduce the effective DOS and the Stoner factor. 
Since T$_C$ in the doped hexaborides
is of the same order, 700 K, it is tempting to assign the observed T$_C$ to the mechanism
of Stoner magnetism due to the B$_6$ vacancy. However,
the typical behavior for ferromagnets is that
the temperature at which S goes below the critical value for ferromagnetism, is generally
much larger than the Curie temperature (T$_C$) for real materials \cite{jan}. 
By taking into account additional
smearing effects on the DOS coming from thermally induced disorder of the structure
one obtains more realistic values of T$_C$ in mean-field calculations \cite{jp}. The
atomic displacements from thermal vibrations at 700 K can be of the order 0.25 a.u. 
(corresponding to 7-8 percent of the B-B distance) for a material 
with a Debye temperature of 350 K, while large smearing effects
on DOS peaks appear already at a lower degree of disorder \cite{fesi}.
The present calculations
include only the T-dependence via the Fermi-Dirac distribution and the effects from thermal disorder
are expected to reduce the moment further. 
Therefore,  it is difficult to understand the high T$_C$ in the real materials from the
T-dependence of the moments within the Stoner mechanism.
One possibility would be that the local structure around a B$_6$ vacancy somehow can escape
effects of thermal disorder, in which case it is justified to consider only the effect 
Fermi-Dirac smearing. The position of $E_F$ relative to the peak appears to be at the optimal
place for the largest possible Stoner factor for this supercell, 
and an increased
height of the peak at larger T would be opposite to the normal T-dependence of the DOS. 

Another mechanism leading to magnetism is due to a gain of Coulomb energy in addition to the exchange
energy, by means of charge transfers. This mechanism
can be activated if there is a large derivative $N'$ of the DOS of an impurity band,
leading to very small moments near the impurity \cite{tj}. 
This difference with the normal
Stoner mechanism, suggests that this weak magnetism could resist better to temperature and broadening
of the DOS by disorder, since $N'$ can be large within a wider energy interval 
compared to that of a peak in the DOS. However, the results from
calculations for a La impurity in a 3$x$3$x$3 cell give only a partial support to this expectation \cite{tj2}.
The DOS of this cell is such that $N'$ is maximal a bit below $E_F$, and even if there is a small
moment at 600 K, it tends to zero
at low T. It is difficult to have a sizable moment both at low
and high T. By optimizing the position of $N'$ relative to $E_F$ (and the amplitude of $N$)
by a combination of the virtual crystal approximation and a non-conventional choice of linearization
energies in the band calculation, it is possible to have a weak moment within some range of T. 
Such band results are not ab-initio, but they are useful to show a correlation of moment and
charge transfers. They also show that magnetism within this mechanism depend on fine details of
the electronic structure near the impurity, which in turn suggest that such magnetism should 
be a result of circumstances. Furthermore, it is expected that thermal disorder should be as important
for this type of magnetism as it is for the standard Stoner magnetism.

In conclusion,
the band results confirm that weak magnetism is possible, either as standard Stoner
magnetism around a B$_6$ vacancy or assisted by Coulomb energies near a La-impurity, 
but it is more difficult to understand the high $T_C$ values
from both of these mechanisms. These results and recent experimental reports \cite{tera,mori}
motivate similar electronic structure studies of the properties near iron impurities in these materials.
However, the exact state of iron contamination, impurity sites or clustering on surfaces, is yet not clear.

\vspace{0.5cm}


\begin{figure}[tb!]
\leavevmode\begin{center}\epsfxsize8.6cm\epsfbox{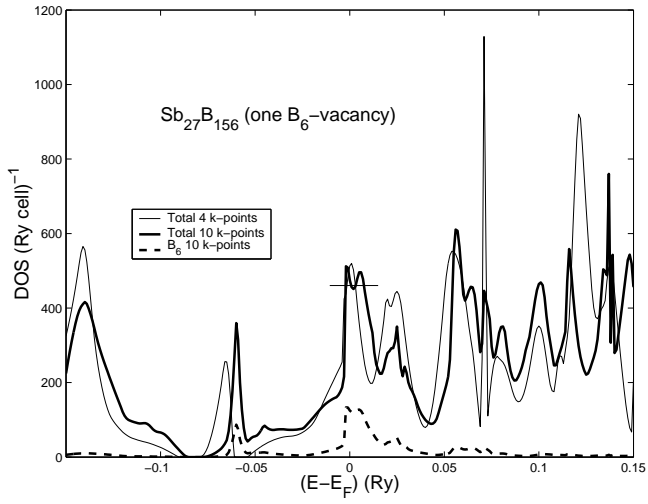}\end{center}
\caption{
Total DOS of Sr$_{27}$B$_{156}$ near $E_F$, using 4 and 10 k-points. The broken line is the
partial DOS on B$_6$ closest to the vacancy. The vertical line indicate the limit for Stoner magnetism.}
\end{figure}

\begin{figure}[tb!]

\leavevmode\begin{center}\epsfxsize8.6cm\epsfbox{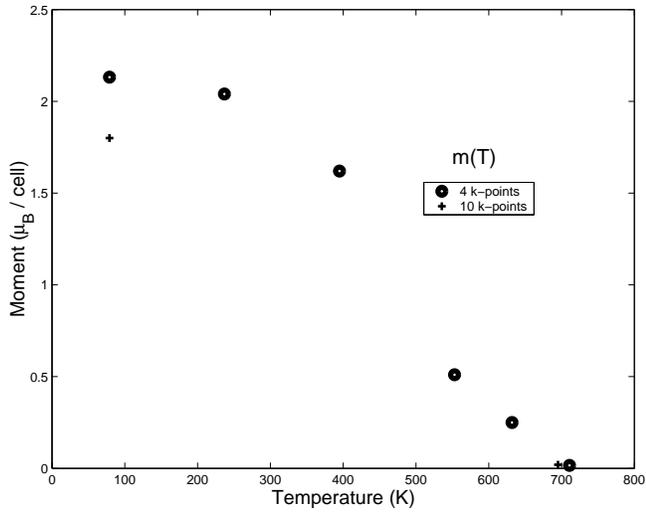}\end{center}
\caption{
 Calculated magnetic moment ($\mu_B$ per cell) as function of temperature for Sr$_{27}$B$_{156}$.
 }
\end{figure}

\end{document}